\documentclass[doublecol]{epl2}
\title{Genuine tripartite entanglement of W state subject to Hawking effect of a Schwarzschild black hole}
\shorttitle{Title} %Insert here a short version of the title if it exceeds 70 characters

\author{Shu-Min Wu$^1$\footnote{Email: smwu@lnnu.edu.cn },  Xiao-Wei Fan$^1$\footnote{Email: xwfan0825@163.com },  Xiao-Li Huang$^1$\footnote{Email: huangxiaoli1982@foxmail.com (corresponding author)}, Hao-Sheng Zeng$^2$\footnote{Email: hszeng@hunnu.edu.cn (corresponding author)}}
\shortauthor{F. Author \etal}

\institute{
  \inst{1} Department of Physics, Liaoning Normal University, Dalian 116029, China\\
  \inst{2} Department of Physics, Hunan Normal University, Changsha 410081, China
}

\abstract{We study the genuine tripartite entanglement (GTE), one-tangle and two-tangle of W state of fermionic fields in the background of a Schwarzschild black hole.
We find that, with the increase of the Hawking temperature, the GTE of W state first decreases and then tends to zero, while the GTE of GHZ state first decreases and then freezes.
We also find that the Hawking effect can completely destroy the two-tangle of W state, whilie one-tangle first decreases and then appears freezing phenomenon with the growth of the Hawking temperature. These results are helpful to guide us to select appropriate quantum states and quantum resources to deal with relativistic quantum information tasks.}

\begin{document}

\maketitle

\section{Introduction}

Quantum entanglement arises from the tensorproduct structure of the
Hilbert space and the superposition principle. Quantum entanglement is one of the most valuable resource of quantum tasks, such as quantum teleportation \cite{L1}, dense coding \cite{L2} and quantum key distribution \cite{L3}. One of the most famous multipartite entangled states is the W state.
Up to now, many experiments can generate W states.
By tracing out a single particle from the W state, the remaining bipartite state is still entangled. Therefore, the W state exhibits a high persistency
of quantum entanglement against particle loss \cite{L4}.
W state also has the advantage that bipartite entanglement of its particles persists even
if the third particle suffers from decoherence \cite{L6,L7}. Because of these merits, W state has been widely used in quantum information \cite{L5}.

General relativity predicts black holes, and the gravitational waves are detected by the Virgo and LIGO detectors from a binary black hole merger system that indirectly proves the existence of black holes in our unverse \cite{L8}. On the other hand, the releasing of the images of M87* \cite{HL25,HL26,HL27,HL28,HL29,HL30} and Sgr A*  \cite{HL31} directly proved the existence of black holes. Black holes are basal objects in gravity of Einstein which are totally composed of several conserved quantities, such as angular momentum, mass, and charge. The interior and exterior of the event horizon of the black holes are separated. The Hawking radiation comes from the particle-antiparticle pairs of autonomous creation by quantum fluctuations in vicinal event horizon \cite{qHL31,qqHL31,qqqHL31}. Nowadays, there are a growing number of researchers who pay attention to the study of black holes. Specifically, the Schwarzschild solution of the four spacetime dimensions may represent the simplest black hole. Quantum information takes a significant role in the study of the information loss problem and thermodynamics of black holes \cite{L9,L10,L11,L12}. Therefore, it is important that the influence of the relativistic effect on quantum maneuverability is studied in curved spacetime. Bipartite entanglement has been studied extensively in Schwarzschild spacetime \cite{HL32,HL33,HL34,HL35,HL36,HL36QW}, while tripartite entanglement of W state has not been studied in Schwarzschild spacetime.
This is the main motivation of our study.

Various types of tripartite entanglement are important and interesting for Alice, Bob and Charlie. Bipartite entanglement includes entanglement between two particles and entanglement between one particle and the remaining two particles as one party. Tripartite entanglement exists in the whole tripartite system, which cannot be simplified by any combination of various bipartite entanglement \cite{L13}. Recently, quantum entanglement has come a long way in quantum information theory. The concepts of one-tangle and two-tangle have been proposed. The one-tangle describes the bipartite entanglement between one particle and the remaining two particles, and the two-tangle describes the bipartite entanglement in any reduced bipartite systems \cite{L14,LAA14}. In terms of one-tangle and two-tangle, we can define the measure of tripartite entanglement, which is called the residual-tangle or residual entanglement  \cite{L14}. We define the smallest residual entanglement as genuine tripartite entanglement (GTE).  GTE is an important type of quantum entanglement which  provides significant advantages in quantum tasks. GTE is a crucial resources for measurement-based quantum computing  \cite{L15} and high-precision metrology \cite{L16}, and has an important function in quantum phase transitions \cite{L17,L18}.

In this paper, we study GTE, one-tangle and two-tangle of W state for free fermionic modes  in the background of eternal Schwarzschild black hole. We assume that Alice, Bob and Charlie initially share a tripartite pure state in an asymptotically flat region. Then Alice still stays stationary at an asymptotically flat region, while Bob and Charlie hover near the event horizon of the black hole. Alice observes a vacuum state, which would be detected as a thermal state from Bob and Charlie's point of view. The Hawking temperature $T$ of the thermal bath rests with the surface gravity $\kappa$ of the black hole from a viewpoint of general relativity. We will study the influence of the Hawking effect on GTE, one-tangle and two-tangle of W state in Schwarzschild spacetime, by making a comparison with the GHZ state in Schwarzschild spacetime.

The structure of the paper is as follows. In the next section, we briefly introduce the measures of one-tangle, two-tangle and GTE. In the third section, we describe the quantization of Dirac fields in a Schwarzschild black hole.  In the fourth section, we study the influence of the Hawking effect on GTE, one-tangle and two-tangle in Schwarzschild spacetime. The last section is devoted to a brief conclusion.

\section{Measures of one-tangle, two-tangle and genuine tripartite entanglement}
Quantifying entanglement in tripartite systems is generally complicated. A method to determine the existence of tripartite correlation in an entangled state is to explore the entanglement distribution in the tripartite system. Different from classical correlations, quantum entanglement is monogamous, which is not freely shared in multiple subsystems of a quantum system \cite{L19}. Therefore, we can use the residual entanglement as a way for measuring nonclassical correlations of the tripartite systems. The basis for analyzing residual entanglement is negativity.
Negativity has been used extensively, which quantifies the entanglement in a state as the degree of seeing whether the system is still entangled. A system is entangled when its density matrix of the partial transpose has negative eigenvalues. The bipartite entanglement of a tripartite system has two types: entanglement between two subsystems, and entanglement between one subsystem and the remaining two subsystems as one party.
The bipartite entanglement between one subsystem and the remaining two subsystems is called one-tangle,
\begin{eqnarray}\label{S1}
N_{\alpha(\beta\gamma)}=\|\rho^{T_{\alpha}}_{\alpha\beta\gamma}\|-1.
\end{eqnarray}
The bipartite entanglement between two subsystems is called two-tangle,
\begin{eqnarray}\label{S2}
N_{\alpha\beta}=\|\rho^{T_{\alpha}}_{\alpha\beta}\|-1.
\end{eqnarray}
Here $T_{\alpha}$ is partial transpose of $\rho_{\alpha\beta\gamma}$ and $\rho_{\alpha\beta}$ relative to observer $\alpha$. Note that $\|A\|-1 $ is actually equal to the two times of the sum of absolute values of the negative eigenvalues of the operator $A$  \cite{L14,LAA14}. Thus, one-tangle and two-tangle also can be expressed as
\begin{eqnarray}\label{S3}
 N_{\alpha(\beta\gamma)}=2\sum^{n}_{i=1}|\lambda^{(-)}_{\alpha(\beta\gamma)}|^{i},
\end{eqnarray}
\begin{eqnarray}\label{S4}
 N_{\alpha\beta}=2\sum^{n}_{j=1}|\lambda^{(-)}_{\alpha\beta}|^{j}.
\end{eqnarray}

For three parties, the Coffman-Kundu-Wootters inequality describes the monogamy constraint
\begin{eqnarray}\label{S5}
 N^{2}_{\alpha(\beta\gamma)}\geq\ N^2_{\alpha\beta}+N^2_{\alpha\gamma}.
\end{eqnarray}
The right hand side of inequality (5) represents the sum of the square of the two two-tangles between subsystem $\alpha$ and the every one of remaining subsystems. The other side quantifies the square of the one-tangle between subsystem $\alpha$ and the remaining subsystems.

We choose the minimum of each non-negative difference between the two sides of inequality in a subsystem, which is called minimally residual tripartite entanglement
\begin{eqnarray}\label{S6}
 E_{(\alpha|\beta|\gamma)}=\min_{(\alpha,\beta,\gamma)}[N^{2}_{\alpha(\beta\gamma)}- N^2_{\alpha\beta}-N^2_{\alpha\gamma}],
\end{eqnarray}
where $(\alpha,\beta,\gamma)$ shows all the permutations of the three mode indices.
In the tripartite quantum system, a significance of the minimally residual entanglement denotes the genuine tripartite entanglement (GTE) shared by the three subsystems \cite{L14,LAA14}.

\section{Quantization of Dirac fields in a Schwarzschild black hole }
The metric in Schwarzschild spacetime can be expressed as
\begin{eqnarray}\label{S7}
ds^2&=&-(1-\frac{2M}{r}) dt^2+(1-\frac{2M}{r})^{-1} dr^2\nonumber\\&&+r^2(d\theta^2
+\sin^2\theta d\varphi^2),
\end{eqnarray}
where $M$ is the mass of the black hole, and $r$ is radial coordinates. We set $G$, $c$, $\hbar$ and $k_B$ as unity in this paper. The Dirac equation \cite{HL46} in a general spacetime is written as
\begin{eqnarray}\label{S8}
[\gamma^a e_a{}^\mu(\partial_\mu+\Gamma_\mu)]\Phi=0,
\end{eqnarray}
where $\gamma^a$ are the Dirac matrices, the four-vectors $e_a{}^\mu$
is the inverse of the tetrad $e^a{}_\mu$ defined by
$g_{\mu\nu}=\eta_{ab}e^{a}{}_{\mu}e^b{}_{\nu}$ with $\eta_{ab}={\rm
diag}(-1, 1, 1, 1)$, and $\Gamma_\mu=
\frac{1}{8}[\gamma^a,\gamma^b]e_a{}^\nu e_{b\nu;\mu}$ are the spin
connection coefficients.

Specifically, the Dirac equation in Schwarzschild spacetime can be expressed as \cite{HL47}
\begin{eqnarray}\label{S9}
&-&\frac{\gamma_{0}}{\sqrt{1-\frac{2M}{r}}}\frac{\partial\Phi}{\partial t}+\gamma_{1}\sqrt{1-\frac{2M}{r}}[\frac{\partial}{\partial r}+\frac{1}{r}\\ \nonumber&+&\frac{M}{2r(r-2M)}]\Phi
+\frac{\gamma_{2}}{r}(\frac{\partial}{\partial\theta}+\frac{\cot\theta}{2})\Phi
+\frac{\gamma_{3}}{r\sin\theta}\frac{\partial\Phi}{\partial\varphi}=0,
\end{eqnarray}
where $\gamma_{i}$ (i = 0, 1, 2, 3) are the Dirac matrices. By solving Eq.(9), we obtain the positive (fermion) frequency outgoing solutions outside and inside regions of the event horizon \cite{HL47,HL48,HL49,HL49Q}
\begin{eqnarray}\label{S10}
\Phi^+_{{k},{\rm out}}\sim \phi(r) e^{-i\omega u},
\end{eqnarray}
\begin{eqnarray}\label{S11}
\Phi^+_{{k},{\rm in}}\sim \phi(r) e^{i\omega u},
\end{eqnarray}
where $\phi(r)$ represents four-component Dirac spinor, $\omega$ is a monochromatic frequency, $k$ is the wave vector, $\omega=|k|$ in the massless Dirac field and $u=t-r_{*}$ with the tortoise coordinate $r_{*}=r+2M\ln\frac{r-2M}{2M}$.
Therefore, we expand Dirac field $\Phi$ through Eqs.(10) and (11) as
\begin{eqnarray}\label{S12}
\Phi&=&\int dk[\hat{a}^{in}_{k}\Phi^+_{{k},{\rm in}}+\hat{b}^{in\dag}_{-k}\Phi^-_{{-k},{\rm in}}\\ \nonumber&+&\hat{a}^{out}_{k}\Phi^+_{{k},{\rm out}}+\hat{b}^{out\dag}_{-k}\Phi^-_{{-k},{\rm out}}],
\end{eqnarray}
where $\hat{a}^{in}_{k}$ and $\hat{b}^{in\dag}_{-k}$ are the annihilation operator of fermion and the creation operator of antifermion for the interior of the event horizon, respectively, and $\hat{a}^{out}_{k}$ and $\hat{b}^{out\dag}_{-k}$ are the annihilation operator of fermion and creation operator of antifermion for the exterior of the event horizon in the Schwarzschild black hole, respectively.

According to Domour and Ruffini's suggestions  \cite{HL50}, one provides a complete basis for the positive energy mode (Kruskal mode) by the analytic extension of Eqs.(10) and (11)
\begin{eqnarray}\label{S13}
\Psi^+_{{k},{\rm out}}=e^{-2\pi M\omega}\Phi^-_{{-k},{\rm in}}+e^{2\pi M\omega}\Phi^+_{{k},{\rm out}},
\end{eqnarray}
\begin{eqnarray}\label{S14}
\Psi^+_{{k},{\rm in}}=e^{-2\pi M\omega}\Phi^-_{{-k},{\rm out}}+e^{2\pi M\omega}\Phi^+_{{k},{\rm in}}.
\end{eqnarray}
Therefore, we also use the Kruskal modes to expand the Dirac field \cite{HL48}
\begin{eqnarray}\label{S15}
\Phi&=&\int dk[2\cosh(4\pi M\omega)]^{-\frac{1}{2}}[\hat{c}^{in}_{k}\Psi^+_{k,in}\\ \nonumber&+&\hat{d}^{in\dagger}_{-k}\Psi^-_{-k,in}+\hat{c}^{out}_{k}\Psi^+_{k,out}+\hat{d}^{out\dagger}_{-k}\Psi^-_{-k,out}],
\end{eqnarray}
where $\hat{c}^{\sigma}_{k}$ and $\hat{d}^{\sigma\dagger}_{k}$ with $\sigma=(in,out)$ are the annihilation operators of fermion and creation operators of antifermion acting on the Kruskal vacuum. Eqs.(12) and (15) are shown that Dirac field can be decomposed by Schwarzschild and Kruskal modes, respectively, which lead to the Bogoliubov transformations between Schwarzschild and Kruskal operators
\begin{eqnarray}\label{S16}
\hat{c}^{out}_{k}=\frac{1}{\sqrt{e^{-8\pi M\omega}+1}}\hat{a}^{out}_{k}-\frac{1}{\sqrt{e^{8\pi M\omega}+1}}\hat{b}^{in\dagger}_{-k},
\end{eqnarray}
\begin{eqnarray}\label{S17}
\hat{c}^{out\dagger}_{k}=\frac{1}{\sqrt{e^{-8\pi M\omega}+1}}\hat{a}^{out\dagger}_{k}-\frac{1}{\sqrt{e^{8\pi M\omega}+1}}\hat{b}^{in}_{-k}.
\end{eqnarray}
According to Bogoliubov transformations, the expressions of the Kruskal vacuum and excited states in the Schwarzschild black hole are written as
\begin{eqnarray}\label{S18}
\nonumber |0\rangle_K&=&\frac{1}{\sqrt{e^{-\frac{\omega}{T}}+1}}|0\rangle_{out} |0\rangle_{in}+\frac{1}{\sqrt{e^{\frac{\omega}{T}}+1}}|1\rangle_{out} |1\rangle_{in},\\
|1\rangle_K&=&|1\rangle_{out} |0\rangle_{in},
\end{eqnarray}
where $T=\frac{1}{8\pi M}$ is the Hawking temperature, ${|n\rangle_{out}}$  and ${|n\rangle_{in}}$ are the number
states for the fermion in the exterior region and the antifermion in the interior region of the event horizon of the black hole.

When an outside observer travels through the Kruskal vacuum, Bob's detector records the number of particles, which can be expressed as
\begin{eqnarray}\label{S19}
N_{F}=_{K}\langle0|\hat{a}^{out\dagger}_{k}\hat{a}^{out}_{k}|0\rangle_{K}=\frac{1}{e^{\frac{\omega}{T}}+1}.
\end{eqnarray}
The equation represents that the observer detects a thermal Fermi-Dirac distribution of the particles in the exterior of the  event horizon of the black hole.
\section{The influence of the Hawking effect on GTE, one-tangle and two-tangle in the Schwarzschild black hole}
We assume that Alice, Bob and Charlie initially stay stationary at an asymptotically flat region and share a W state
\begin{eqnarray}\label{S20}
|W\rangle=\frac{1}{\sqrt{3}}[|0_{A}0_{B}1_{C}\rangle+|0_{A}1_{B}0_{C}\rangle+|1_{A }0_{B}0_{C}\rangle],
\end{eqnarray}
where the subscripts $A$, $B$ and $C$ denote the qubits shared by Alice, Bob and Charlie, respectively.
Subsequently, we consider Alice still stays stationary at an asymptotically flat region, while Bob and Charlie hover near the event horizon of the black hole.
According to Eqs.(\ref{S18}) and (\ref{S20}), the wave function of W state can be rewritten as
\begin{eqnarray}\label{SS21}
|\bar W\rangle&=&\frac{1}{\sqrt{3}}[\mu|0_{A}0_{B}0_{\bar B}1_{C}0_{\bar C}\rangle
+\mu|0_{A}1_{B}0_{\bar B}0_{C}0_{\bar C}\rangle
\nonumber\\
&+&\nu|0_{A}1_{B}0_{\bar B}1_{C}1_{\bar C}\rangle+\nu|0_{A}1_{B}1_{\bar B}1_{C}0_{\bar C}\rangle\nonumber\\
&+&\mu^2|1_{A}0_{B}0_{\bar B}0_{C}0_{\bar C}\rangle+\mu\nu|1_{A}0_{B}0_{\bar B}1_{C}1_{\bar C}\rangle\nonumber\\
&+&\mu\nu|1_{A}1_{B}1_{\bar B}0_{C}0_{\bar C}\rangle+\nu^2|1_{A}1_{B}1_{\bar B}1_{C}1_{\bar C}\rangle],
\end{eqnarray}
where $\mu=\frac{1}{\sqrt{e^{-\frac{\omega}{T}}+1}}$ and $\nu=\frac{1}{\sqrt{e^{\frac{\omega}{T}}+1}}$.
Since Bob and Charlie cannot access the modes inside event horizon of the black hole, we should trace over the inaccessible $\bar B$ and $\bar C$ modes. Therefore, by tracing
over the inaccessible modes, we obtain the density matrix

\begin{eqnarray}\label{S22}
 \rho_{ABC}=\frac{1}{3} \left(\!\!\begin{array}{cccccccc}
0 & 0 & 0 & 0 & 0 & 0 & 0 & 0\\
0 & \mu^2 & \mu^2 & 0 & \mu^3 & 0 & 0 & 0\\
0 & \mu^2 & \mu^2 & 0 & \mu^3 &  0 &0  & 0\\
0 & 0 & 0 & 2\nu^2 & 0 & \mu\nu^2 & \mu\nu^2 & 0 \\
0 & \mu^3 & \mu^3 & 0 & \mu^4 &0 & 0 &0 \\
0 & 0 & 0 & \mu\nu^2 & 0 & \mu^2\nu^2 & 0 & 0\\
0 & 0 & 0 & \mu\nu^2 & 0 & 0 & \mu^2\nu^2 &0 \\
0 & 0 & 0 & 0 & 0 & 0 & 0 & \nu^4
\end{array}\!\!\right).
\end{eqnarray}

According to Eq.(\ref{S6}), the GTE of W state in Schwarzschild spacetime can be expressed as
\begin{eqnarray}\label{S23}
E_{(A|B|C)}^W&=&\frac{1}{576}\{8\mu^4-8\mu^2+2\sqrt{2}[35\mu^4+28\mu^4(2\mu^2-1)\nonumber\\
&+&\mu^4(8\mu^4-8\mu^2+1)]^\frac{1}{2}\}^2-\frac{1}{18}\{\sqrt{2}[5\nonumber\\
&+&4(2\mu^2-1)+(8\mu^4-8\mu^2+1)]^\frac{1}{2}-2\}^2.
\end{eqnarray}
From Eq.(\ref{S23}) we can see that the GTE  of W state depends on the Hawking temperature $T$, which means that the Hawking radiation will affect the GTE in the Schwarzschild black hole.
On the other hand, the GTE of GHZ state in curved spacetime reads $E_{(A|B|C)}^{GHZ}=\frac{1}{4}[\mu^2-\mu^2\nu^2+\mu\sqrt{\nu^4\mu^2+\mu^2}]^2$  \cite{HL47}.

\begin{figure}%[ht]
\includegraphics[scale=0.9]{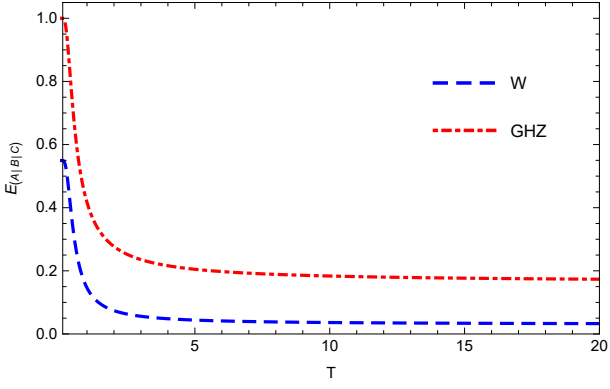}
\caption{The GTE of W and GHZ states as a function of the Hawking temperature $T$.
}\label{Fig1}
\end{figure}

In Fig.\ref{Fig1}, we plot the GTE of W and GHZ states as a function of the Hawking temperature $T$ in the Schwarzschild black hole.
We find that the GTE of W state first decreases and then tends to zero with the increase of the Hawking temperature $T$, while GTE of GHZ state first decreases and then freezes with the increase of the Hawking temperature $T$ \cite{HL47}. We also find that the GTE of W state is smaller than that  of GHZ state.
This implies that the GTE of GHZ state is more effective for resisting the Hawking effect and is more suitable for processing relativistic quantum information.

We also study bipartite entanglement $N_{A(BC)}$, $N_{B(AC)}$, $N_{C(AB)}$, $N_{AB}$, $N_{AC}$ and $N_{BC}$ for the W state in Schwarzschild spacetime.
Firstly, we consider the one-tangles $N_{A(BC)}$, $N_{B(AC)}$ and $N_{C(AB)}$.
Taking the transpose of $\rho_{ABC}$ with respect mode $A$, we get
\begin{eqnarray}\label{S24}
 \rho^{T_{A}}_{ABC}=\frac{1}{3} \left(\!\!\begin{array}{cccccccc}
0 & 0 & 0 & 0 & 0 & \mu^3 & \mu^3 & 0\\
0 & \mu^2 & \mu^2 & 0 & 0 & 0 & 0 & \mu\nu^2\\
0 & \mu^2 & \mu^2 & 0 & 0 &  0 &0  & \mu\nu^2\\
0 & 0 & 0 & 2\nu^2 & 0 & 0 & 0 & 0 \\
0 & 0 & 0 & 0 & \mu^4 &0 & 0 &0 \\
\mu^3 & 0 & 0 & 0 & 0 & \mu^2\nu^2 & 0 & 0\\
\mu^3 & 0 & 0 & 0 & 0 & 0 & \mu^2\nu^2 &0 \\
0 & \mu\nu^2 & \mu\nu^2 & 0 & 0 & 0 & 0 & \nu^4
\end{array}\!\!\right), \nonumber
\end{eqnarray}
which has the negative eigenvalue $\frac{1}{48}[-8\mu^4+8\mu^2-2\sqrt{2}\sqrt{35\mu^4+28\mu^4(2\mu^2-1)+\mu^4(8\mu^4-8\mu^2+1)}]$.
Thus the one-tangle $N_{A(BC)}$ is
\begin{eqnarray}\label{S25}
 N_{A(BC)}&=&\frac{1}{24}\{8\mu^4-8\mu^2+2\sqrt{2}[35\mu^4+28\mu^4\nonumber\\
&&(2\mu^2-1)+\mu^4(8\mu^4-8\mu^2+1)]^\frac{1}{2}\}.
\end{eqnarray}
Similarly, taking the transpose with respect the mode $B$,
\begin{eqnarray}\label{S26}
 \rho^{T_{B}}_{ABC}=\frac{1}{3} \left(\!\!\begin{array}{cccccccc}
0 & 0 & 0 & \mu^2 & 0 & 0 & \mu^3 & 0\\
0 & \mu^2 & 0 & 0 & \mu^3 & 0 & 0 & \mu\nu^2\\
0 & 0 & \mu^2 & 0 & 0 & 0 & 0 & 0\\
\mu^2 & 0 & 0 & 2\nu^2 & 0 & 0 & \mu\nu^2 & 0 \\
0 & \mu^3 & 0 & 0 & \mu^4 &0 & 0 &0 \\
0 & 0 & 0 & 0 & 0 & \mu^2\nu^2 & 0 & 0\\
\mu^3 & 0 & 0 & \mu\nu^2 & 0 & 0 & \mu^2\nu^2 &0 \\
0 & \mu\nu^2 & 0 & 0 & 0 & 0 & 0 & \nu^4
\end{array}\!\!\right). \nonumber
\end{eqnarray}
Due to the complexity of the expression of $N_{B(AC)}$, we do not write it out here. Since Bob and Charlie are symmetric, we obtain $N_{B(AC)}=N_{C(AB)}$.
\begin{figure}%[ht]
\includegraphics[scale=0.9]{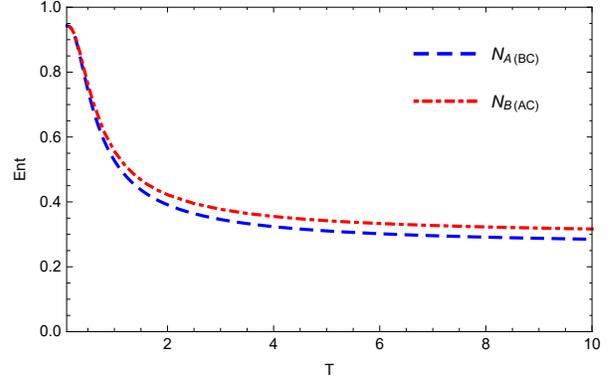}
\caption{The one-tangles of W state as a function of the Hawking temperature $T$.
}\label{Fig2}
\end{figure}

In Fig.\ref{Fig2}, we plot the one-tangles of W state as a function of the Hawking temperature $T$. We can see that one-tangles of W state first decrease and then appear freezing phenomenon with the increase of the Hawking temperature $T$. We also find that one-tangle of GHZ state is bigger than one-tangle of W state in curved spacetime, which means that one-tangle of GHZ state can effectively resist Hawking effect \cite{HL47}. This indicates that one-tangle of GHZ state is more suitable for processing relativistic quantum information.
It has been shown that\cite{HL51,HL52}, however, the coherence of W state is always bigger than the coherence of GHZ state in Schwarzschild spacetime, meaning that the coherence of W state is more suitable for processing relativistic quantum information than the GHZ state.
Therefore, we should choose suitable quantum resources as required to process relativistic quantum information.

Next, we consider the two-tangles $N_{AB}$, $N_{AC}$ and $N_{BC}$.
Taking trace over the mode $C$ from $\rho_{ABC}$, one gets
\begin{eqnarray}\label{S27}
\rho_{AB}=\frac{1}{3} \left(\!\!\begin{array}{cccc}
\mu^2 &0 & 0 &0 \\
0 & 1+\nu^2 & \mu & 0 \\
0 & \mu & \mu^2 & 0 \\
0 & 0 & 0 &\nu^2
\end{array}\!\!\right).
\end{eqnarray}
The transpose with respect mode $A$ is
\begin{eqnarray}\label{S28}
\rho^{T_{A}}_{AB}=\frac{1}{3} \left(\!\!\begin{array}{cccc}
\mu^2 &0 & 0 &\mu \\
0 & 1+\nu^2 & 0 & 0\\
0 & 0 & \mu^2 &0 \\
\mu & 0 & 0 &\nu^2
\end{array}\!\!\right).
\end{eqnarray}
According to Eqs.(\ref{S4}) and (\ref{S28}), the two-tangle $N_{AB}$ can be expressed as
\begin{eqnarray}\label{S31}
 N_{AB}=\frac{1}{6}[\sqrt{2}\sqrt{4(2\mu^2-1)+(8\mu^4-8\mu^2+1)+5}-2].
\end{eqnarray}
Since Bob and Charlie are symmetric, we have $N_{AB}$=$N_{AC}$.

Similarly, we can get
\begin{eqnarray}\label{S29}
\rho_{BC}=\frac{1}{3} \left(\!\!\begin{array}{cccc}
\mu^4 &0 & 0 & 0 \\
0 & \mu^2(1+\nu^2) & \mu^2 & 0\\
0 & \mu^2 & \mu^2(1+\nu^2) &0 \\
0 & 0 & 0 &\nu^2(2+\nu^2)
\end{array}\!\!\right),
\end{eqnarray}
and its transpose
\begin{eqnarray}\label{S30}
\rho^{T_{B}}_{BC}=\frac{1}{3} \left(\!\!\begin{array}{cccc}
\mu^4 &0 & 0 &\mu^2 \\
0 & \mu^2(1+\nu^2) & 0 & 0\\
0 & 0 & \mu^2(1+\nu^2) &0 \\
\mu^2 & 0 & 0 &\nu^2(2+\nu^2)
\end{array}\!\!\right).
\end{eqnarray}
Employing  Eqs.(\ref{S4}) and (\ref{S30}), the two-tangle $N_{BC}$ can be written as
\begin{eqnarray}\label{S32}
 N_{BC}&=&\frac{1}{12}(-12-8\mu^4+16\mu^2 \\ \nonumber
 &+& 2\sqrt{2}\sqrt{18+40\mu^4-48\mu^2}).
\end{eqnarray}

\begin{figure}%[ht]
\includegraphics[scale=0.9]{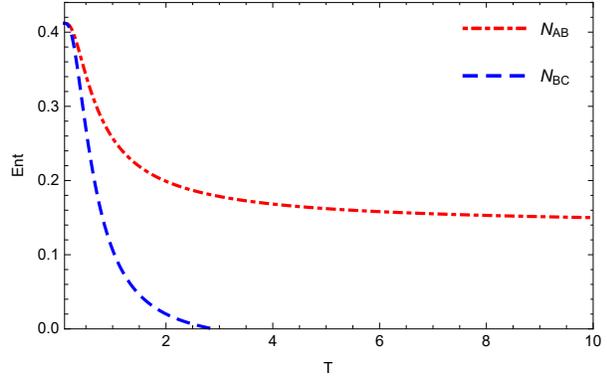}
\caption{The $N_{AB}$ and $N_{BC}$ of W state as a function of the Hawking temperature $T$.
}\label{Fig3}
\end{figure}

In Fig.\ref{Fig3}, we plot the two-tangles of W state as a function of the Hawking temperature $T$. It is shown that the two-tangle $N_{AB}$ between Alice and Bob first decreases and then freezes with the increase of the Hawking temperature $T$. However, the two-tangle $N_{BC}$ between Bob and Charlie first decreases and then suffers from sudden death with the growth of the Hawking temperature $T$. This means that the Hawking effect completely destroys the two-tangle $N_{BC}$ of W state. These results are in contrast with the two-tangles of the tripartite GHZ state, which are zero in curved spacetime \cite{HL47}.

Finally, we compare fermionic entanglement with bosonic entanglement of tripartite states in Schwarzschild spacetime. We initially assume that Alice, Bob and Charlie stay stationary at an asymptotically flat region and share GHZ and W states of bosonic field. According to Bogoliubov transformations, the Kruskal vacuum and excited states of bosonic field in Schwarzschild spacetime can be expressed as
\begin{eqnarray}\label{S31}
|0\rangle^B_K&=&\sqrt{1-e^{-\frac{\omega}{T}}}\sum^\infty_{n=0}e^{-\frac{n\omega}{2T}}|n\rangle^B_{out} |n\rangle^B_{in},\\
\nonumber
|1\rangle^B_K&=&(1-e^{-\frac{\omega}{T}})\sum^\infty_{n=0}e^{-\frac{n\omega}{2T}}\sqrt{n+1}|n+1\rangle^B_{out} |n\rangle^B_{in},
\end{eqnarray}
where ${|n\rangle^B_{out}}$  and ${|n\rangle^B_{in}}$ are the number
states for the boson in the exterior region and the antiboson in the interior region of the event horizon of the black hole, respectively \cite{HL33}. Hereafter, we omit the mark $B$ for simplicity unless it causes confusion.
Because the tripartite entanglement of bosonic field is very complex in Schwarzschild spacetime, we consider a simpler model:  Charlie hovers near the event horizon of the black hole, while Alice and Bob still stay stationary at an asymptotically flat region. According to Eq.(\ref{S31}), the wave functions of GHZ and W states can be rewritten as
\begin{eqnarray}\label{S32}
|GHZ\rangle^{B}_{ABC}&=&\frac{1}{\sqrt{2}\alpha}\sum^\infty_{n=0}\gamma^n(|00n\rangle|n\rangle_{in} \\ \nonumber
&&+\frac{\sqrt{n+1}}{\alpha}|11n+1\rangle|n\rangle_{in}),
\end{eqnarray}
\begin{eqnarray}\label{S33}
|W\rangle^{B}_{ABC}&=&\frac{1}{\sqrt{3}\alpha}\sum^\infty_{n=0}\gamma^n(\frac{\sqrt{n+1}}{\alpha}|00n+1\rangle \\ \nonumber
&&+|01n\rangle+|10n\rangle)\bigotimes|n\rangle_{in},
\end{eqnarray}
where $\alpha=\frac{1}{\sqrt{1-e^{-\frac{\omega}{T}}}}$ and $\gamma=\frac{1}{\sqrt{e^{\frac{\omega}{T}}}}$. Since Charlie cannot access the modes inside the event horizon of the black hole, we should trace over the inaccessible mode and obtain the density operators
\begin{eqnarray}\label{S34}
\rho^{B}_{GHZ}&=&\frac{1}{2\alpha^2}\sum^\infty_{n=0}\gamma^{2n}\{|00n\rangle\langle00n| \\ \nonumber
&&+\frac{\sqrt{n+1}}{\alpha}[|00n\rangle\langle11n+1|+|11n+1\rangle\langle00n|] \\ \nonumber &&+\frac{n+1}{\alpha^2}|11n+1\rangle\langle11n+1|\},
\end{eqnarray}
\begin{eqnarray}\label{S35}
\rho^{B}_{W}&=&\frac{1}{3\alpha^2}\sum^\infty_{n=0}\gamma^{2n}\{\frac{n+1}{\alpha^2}|00n+1\rangle\langle00n+1| \\ \nonumber
&&+|01n\rangle\langle01n|+|10n\rangle\langle10n| +\frac{\sqrt{n+1}}{\alpha}[|00n+1\rangle\\ \nonumber &&\langle01n|+|01n\rangle\langle00n+1|+|00n+1\rangle\langle10n|+\\ \nonumber &&|10n\rangle\langle00n+1|] +|01n\rangle\langle10n|+|10n\rangle\langle01n|\}.
\end{eqnarray}
Employing Eqs.(\ref{S3}), (\ref{S4}) and (\ref{S6}), we can obtain the GTE and bipartite entanglement $N^B_{C(AB)}$ of the GHZ and W states of bosonic field in Schwarzschild spacetime. Since the expressions are very complex, we do not write them.
\begin{figure}
\includegraphics[width=3.3in,height=6.0cm]{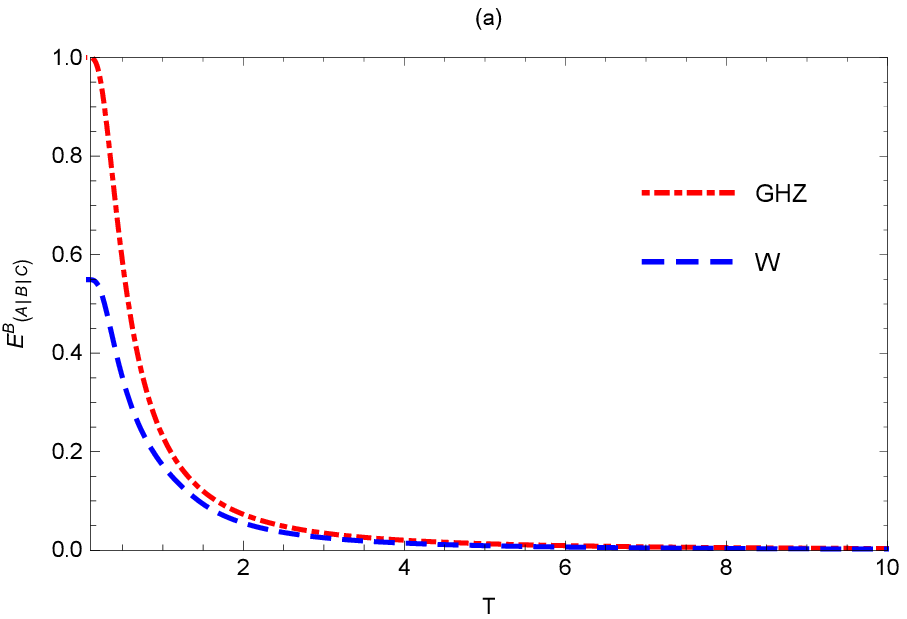}
\label{fig:side:a}
\vspace{0.2cm}
\includegraphics[width=3.3in,height=6.0cm]{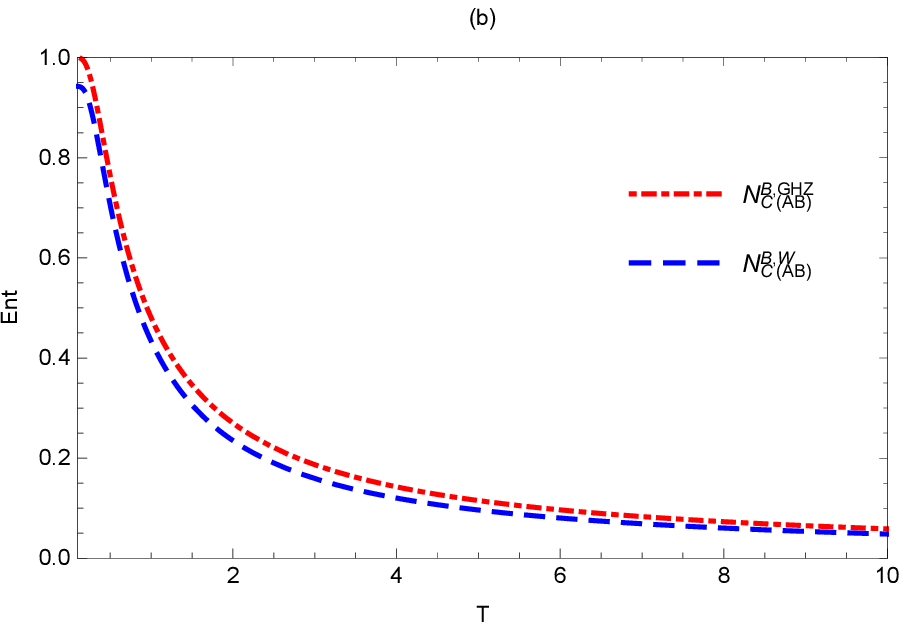}
\label{fig:side:a}
\caption{The GTE, $N^{B,GHZ}_{C(AB)}$ and $N^{B,W}_{C(AB)}$ of GHZ and W states of bosonic field as a function of the Hawking temperature $T$.
 }
\label{Fig4}
\end{figure}

In Fig.\ref{Fig4}, we plot the GTE, $N^{B,GHZ}_{C(AB)}$ and $N^{B,W}_{C(AB)}$ of GHZ and W states of bosonic field as a function of the Hawking temperature $T$.  From Fig.\ref{Fig4} (a), we can see that the GTEs of GHZ and W states of bosonic field vanish in the infinite Hawking temperature $T$, while the GTE of GHZ state of fermionic field always survives in curved spacetime \cite{HL47}. This means that the GTE of GHZ state of fermionic field  may be more suitable for relativistic quantum information tasks.
By the numerical calculation, we find
\begin{eqnarray}\label{S36}
\lim\limits_{T\to\infty}N^{B,GHZ}_{C(AB)}=0,\qquad  \lim\limits_{T\to\infty}N^{B,W}_{C(AB)}=0.
\end{eqnarray}
It means that $N^{B,GHZ}_{C(AB)}$ and $N^{B,W}_{C(AB)}$ of GHZ and W states of bosonic field vanish in the infinite Hawking temperature $T$. However, one-tangles of GHZ and W states of fermionic field  always survive when only Charlie hovers near the event horizon of the black hole \cite{HL47}.  The disparity between the Dirac and scalar fields is caused by the differences between Bose-Einstein and Fermi-Dirac statistics.
This is because that Fermi-Dirac distribution protects tripartite entanglement of fermionic field.  In addition, one-tangles  $N^{B,GHZ}_{A(BC)}$ ($N^{B,GHZ}_{B(AC)}$ ) and $N^{B,W}_{A(BC)}$ ($N^{B,W}_{B(AC)}$ ) of GHZ and W states of bosonic field  can survive for any $T$ \cite{HL520}. This conclusion is consistent with the fermionic field \cite{HL47}.

%------------------------------------------------------------------------------------------------------------------------------------------------------------------------------------------------%
\section{ Conclusions  \label{GSCDGE}}
%--------------------------------------------------------------------------------
The effect of the Hawking effect on the genuine tripartite entanglement (GTE), one-tangle and two-tangle of W state in Schwarzschild spacetime have been investigated. We assume that Alice, Bob and Charlie initially share a W state  at an asymptotically flat region.
Then Alice still stays stationary at an asymptotically flat region, while Bob and Charlie hover near the event horizon of the black hole.
We have found that, with the increase of the Hawking temperature, the GTE of W state first decreases and then approaches zero, while GTE of GHZ state first decreases and then appears freezing phenomenon \cite{HL47}. This implies that the GTE of GHZ state is more effective for resisting Hawking effect.

We have also found that, with the growth of the Hawking temperature, the two-tangle between Alice and Bob (Charlie) of W state first decreases and then freezes, while two-tangle between Bob and Charlie first reduces and then suffers from sudden death. This is different from the case of GHZ state, whose two-tangles are zero in Schwarzschild spacetime\cite{HL47}.
We have shown that the one-tangle of W state first decreases and then appears freezing phenomenon with the increase of the Hawking temperature, which is always smaller than the one-tangle of GHZ state in curved spacetime. This is different from the behavior of quantum coherence, where the coherence of W state is bigger than that of GHZ state in Schwarzschild spacetime \cite{HL51,HL52}. Therefore, for different quantum states, we should choose suitable quantum resources to process relativistic quantum information.

Finally, we compare bosonic entanglement with fermionic entanglement of tripartite states when only Charlie hovers near the event horizon of the Schwarzschild black hole.
We find that the GTEs of GHZ and W states of bosonic field reduce to zero with the growth of the Hawking temperature, while the GTE of GHZ state of fermionic field
can survive for any Hawking temperature. We also find that not all  the one-tangles of GHZ and W states of bosonic field can survive  in the infinite Hawking temperature limit, while  all  one-tangles of GHZ and W states of fermionic field always survive in Schwarzschild spacetime.
This is because that Fermi-Dirac distribution protects tripartite entanglement of fermionic field in Schwarzschild spacetime. It means that tripartite entanglement of fermionic field is more suitable for processing relativistic quantum information.

\acknowledgments
This work is supported by the National Natural
Science Foundation of China (Grant Nos. 12205133, 1217050862, 11275064, 11975064 and 12075050 ), LJKQZ20222315 and 2021BSL013.

\end{document}